\documentclass[11pt,a4paper]{article}
\usepackage[hyperref]{emnlp2018}
\usepackage{times}
\usepackage{latexsym}
\usepackage{arydshln}
\usepackage{url}
\usepackage{graphicx}
\usepackage{comment}
\usepackage{xspace}

\aclfinalcopy 

\newcommand{\mlp}{\textsc{mlp}\xspace}
\newcommand{\lstm}{\textsc{lstm}\xspace}
\newcommand{\bilstm}{\textsc{bilstm}\xspace}

\newcommand{\cnn}{\textsc{cnn}\xspace}
\newcommand{\tcn}{\textsc{tcn}\xspace}
\newcommand{\ir}{\textsc{ir}\xspace}
\newcommand{\ml}{\textsc{ml}\xspace}
\newcommand{\nlp}{\textsc{nlp}\xspace}

\newcommand{\bioasq}{BioASQ\xspace}

\newcommand{\aueb}{AUEB\xspace}
\newcommand{\auebnlp}{\textsc{aueb-nlp}\xspace}

\newcommand{\drmm}{\textsc{drmm}\xspace}
\newcommand{\pacrr}{\textsc{pacrr}\xspace}
\newcommand{\termpacrr}{\textsc{term-pacrr}\xspace}
\newcommand{\pacrrdrmm}{\textsc{pacrr-drmm}\xspace}
\newcommand{\abeldrmm}{\textsc{abel-drmm}\xspace}

\newcommand{\bmtf}{\textsc{bm25}\xspace}
\newcommand{\idf}{\textsc{idf}\xspace}
\newcommand{\map}{\textsc{map}\xspace}
\newcommand{\gmap}{\textsc{gmap}\xspace}
\newcommand{\medline}{\textsc{Medline}\xspace}
\newcommand{\relu}{\textsc{relu}\xspace}
\newcommand{\bcnn}{\textsc{bcnn}\xspace}
\newcommand{\abcnn}{\textsc{abcnn}\xspace}

\title{\aueb at BioASQ 6: Document and Snippet Retrieval}

\author{Georgios-Ioannis Brokos$^1$\textnormal{,} 
Polyvios Liosis$^1$\textnormal{,} 
Ryan McDonald$^{1,3}$\textnormal{,} \\
\textbf{Dimitris Pappas}$^{1,2}$ \and \textbf{Ion Androutsopoulos}$^1$\\
\\
$^1$Dept. of Informatics, Athens University of Economics and Business, Greece\\
$^2$Institute for Language and Speech Processing, Research Center `Athena', Greece\\
$^3$Google AI
}

\date{}

\begin{document}
\maketitle

\begin{abstract}
We present \aueb's submissions to the \bioasq 6 document and snippet retrieval tasks (parts of Task 6b, Phase A). Our models use novel extensions to deep learning architectures that operate solely over the text of the query and candidate document/snippets. Our systems scored at the top or near the top for all batches of the challenge, highlighting the effectiveness of deep learning for these tasks.
\end{abstract}


\section{Introduction}

\bioasq \cite{tsatsaronis2015overview} is a biomedical document classification, document retrieval, and question answering competition, currently in its sixth year.\footnote{Consult \url{http://bioasq.org/}.} We provide an overview of \aueb's submissions to the document and snippet retrieval tasks (parts of Task 6b, Phase A) of \bioasq 6.\footnote{For further information on the \bioasq 6 tasks, see \url{http://bioasq.org/participate/challenges}.} In these tasks, systems are provided with English biomedical questions and are required to retrieve relevant documents and document snippets from a collection of \medline /PubMed articles.\footnote{\url{http://www.ncbi.nlm.nih.gov/pubmed/}.}

We used deep learning models for both document and snippet retrieval. 
For document retrieval, we focus on extensions to the Position-Aware Convolutional Recurrent
Relevance (\pacrr) model of \newcite{hui2017pacrr} and, mostly, the Deep Relevance Matching Model (\drmm) of \newcite{guo2016deep}, whereas for snippet retrieval we based our work on the Basic Bi-CNN (\bcnn) model of \newcite{YinW2016bcnn}.
Little task-specific pre-processing is employed and the models operate solely over the text of the query and candidate document/snippets.

Overall, our systems scored at the top or near the top for all batches of the challenge. In previous years of the \bioasq challenge, the top scoring systems used primarily traditional \ir techniques \cite{jin2017multi}. Thus, our work highlights that end-to-end deep learning models are an effective approach for retrieval in the biomedical domain.

\section{Document Retrieval} 
\label{sec:documentRetrieval}

For document retrieval, we investigate new deep learning architectures focusing on \emph{term-based interaction models}, where query terms (\emph{q-terms} for brevity) are scored relative to a document's terms (\emph{d-terms}) and their scores are aggregated to produce a relevance score for the document. All models use pre-trained embeddings for all q-terms and d-terms. Details on data resources and data pre-processing are given in Section~\ref{sec:resources}.

\subsection{PACRR-based Models} 
\label{sec:pacrr-bioasq}

\begin{figure}[t]
\centering
\includegraphics[width=3in]{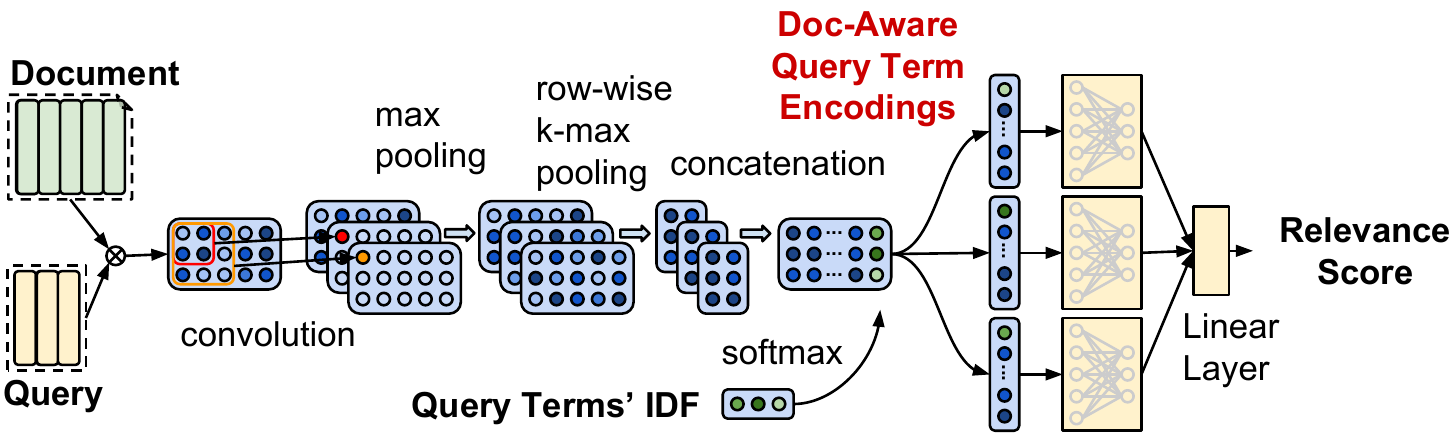}
\caption{\pacrr \cite{hui2017pacrr} and \termpacrr. In \pacrr, an \mlp is applied to the concatenation of the document-aware q-term encodings to produce the relevance score. In \termpacrr, the \mlp is applied \emph{separately} to each document-aware q-term encoding; the resulting scores are combined by a linear layer. }
\label{fig:pacrr}
\end{figure}

The first model we investigate is \pacrr \cite{hui2017pacrr}. In this model, a query-document term similarity matrix $\textit{sim}$ is first computed (Fig.~\ref{fig:pacrr}, left). Each cell $(i, j)$ of $\textit{sim}$ contains the cosine similarity between the embeddings of a q-term $q_i$ and a d-term $d_j$. To keep the dimensions $l_q \times l_d$ of $\textit{sim}$ fixed across queries and documents of varying lengths, queries are padded to the maximum number of q-terms $l_q$, and only the first $l_d$ terms per document are retained.\footnote{We use \textsc{pacrr}-\textit{firstk}, which 
\newcite{hui2017pacrr} recommend when documents fit in memory, as in our experiments.} Then, convolutions of different kernel sizes $n\times n$ ($n=2,\dots,l_g$) are applied to $\textit{sim}$ to capture $n$-gram query-document similarities. For each size  $n\times n$, multiple kernels (filters) are used. Max pooling is then applied along the dimension of the filters (max value of all filters of the same size), followed by $k$-max pooling along the dimension of d-terms to capture the strongest $k$ signals between each q-term and all the d-terms. The resulting matrices (one per kernel size) are concatenated into a single matrix where each row is a document-aware q-term encoding (Fig.~\ref{fig:pacrr}); the \idf of the q-term is also appended, normalized by applying a softmax across the \idf{s} of all the q-terms. Following \newcite{hui2018copacrr}, we concatenate the rows of the resulting matrix into a single vector, which is passed to an \mlp that produces a query-document relevance score.\footnote{\newcite{hui2017pacrr} used an additional \lstm, which was later replaced by the final concatenation \cite{hui2018copacrr}.}

Instead of using an \mlp to score the \emph{concatenation} of all the (document-aware) q-term encodings, a simple extension we found effective was to use an \mlp to \emph{independently} score each q-term encoding (the same \mlp for all q-terms, Fig.~\ref{fig:pacrr}); the resulting scores are aggregated via a linear layer. This version, \termpacrr, performs better than \pacrr, using the same number of hidden layers in the \mlp{s}. Likely this is due to the fewer parameters of \termpacrr's \mlp, which is shared across the q-term representations and operates on shorter input vectors. Indeed, in our early experiments \termpacrr was less prone to over-fitting.\footnote{In the related publication of \newcite{mcdonald2018emnlp} \termpacrr is identical to the \pacrrdrmm model.}

\subsection{DRMM-based Models}
\label{sec:drmm-bioasq}

\begin{figure}[t] 
\includegraphics[width=3in]{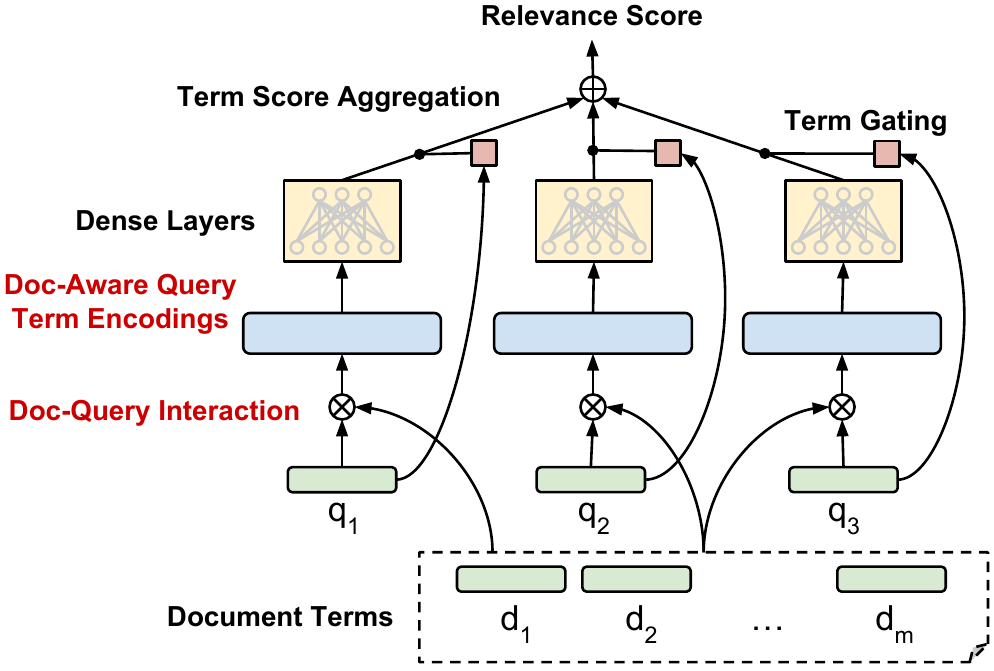}
\caption{Illustration of \drmm \cite{guo2016deep} for three q-terms and $m$ d-terms. The $\otimes$ nodes produce (bucketed) cosine similarity histograms, each capturing the similarity between a q-term and all the d-terms.}
\label{fig:top-level-drmm}
\end{figure}

The second model we investigate is \drmm \cite{guo2016deep} (Fig.~\ref{fig:top-level-drmm}). The original \drmm uses pre-trained word embeddings for q-terms and d-terms, and (bucketed) cosine similarity histograms (outputs of $\otimes$ nodes in Fig.~\ref{fig:top-level-drmm}). Each histogram captures the similarity of a q-term to all the d-terms of a particular document. The histograms, which in this model are the document-aware q-term encodings, are fed to an \mlp (dense layers of Fig.~\ref{fig:top-level-drmm})
that produces the (document-aware) score of each q-term. Each q-term score is then weighted using a gating mechanism (topmost box nodes in Fig.~\ref{fig:top-level-drmm}) that examines properties of the q-term to assess its importance for ranking (e.g., common words are less important). The sum of the weighted q-term scores is the relevance score of the document.

For gating (topmost box nodes of Fig.~\ref{fig:top-level-drmm}), \newcite{guo2016deep} use a linear self-attention:
\[g_i = \textrm{softmax}\Big(w_g^T \, \phi_g(q_i); q_1, \dots, q_n\Big)\]
$\phi_g(q_i)$ is the embedding $e(q_i)$ of the $i$-th q-term, 
or its \idf, $\textrm{idf}(q_i)$; $w_g$ is a weights vector. We found that  $\phi_g(q_i) = [e(q_i); \textrm{idf}(q_i)]$, where `;' is concatenation, was optimal for all \drmm-based models.

\subsubsection{ABEL-DRMM}

The original \drmm \cite{guo2016deep} has two shortcomings. The first one is that it ignores entirely the contexts where the terms occur, in contrast to position-aware models such as \pacrr (Section~\ref{sec:pacrr-bioasq}) or those based on recurrent representations \cite{palangi2016deep}. Secondly, the histogram representation for document-aware q-term encodings is not differentiable, so it is not possible to train the network end-to-end, if one wished to backpropagate all the way to word embeddings.

\begin{figure}[t]
\includegraphics[width=3in]{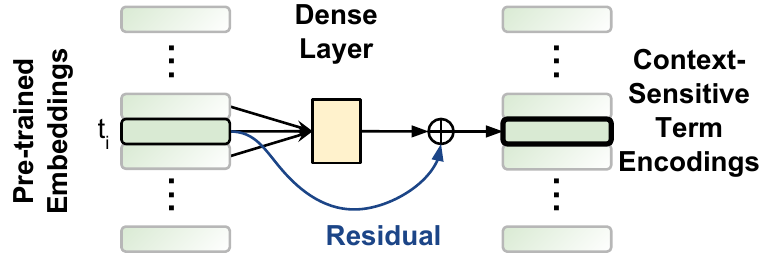}
\caption{Producing \emph{context-sensitive term encodings}.}
\label{fig:ngram-term-reps}
\end{figure}

To address the first shortcoming, we add an encoder (Fig.~\ref{fig:ngram-term-reps}) to produce the \emph{context-sensitive encoding} of each q-term or d-term from the pre-trained embeddings of the previous, current, and next term in a particular query or document. A single dense layer with residuals is used, in effect a one-layer Temporal Convolutional Network (\tcn) \cite{bai2018empirical} without pooling or dilation. The number of convolutional filters equals the dimensions of the pre-trained embedding, for residuals to be summed without transformation.

Specifically, let $e(t_i)$ be the pre-trained embedding for a q-term or d-term term $t_i$. We compute the context-sensitive encoding of $t_i$ as:
\begin{equation}
c(t_i) = \varphi\Big(W_c \, \phi_c(t_i) + b_c\Big) + e(t_i)
\label{eq:ct}
\end{equation}
$W_c$ and $b_c$ are the weights matrix and bias vector of the dense layer, $\varphi$ is the activation function, $\phi_c(t_i) = [e(t_{i-1}); e(t_i); e(t_{i+1})]$, $t_{i-1}, t_{i+1}$ are the tokens surrounding $t_i$ in the query or document. This is an orthogonal way to incorporate context into the model relative to \pacrr. \pacrr creates a query-document similarity matrix and computes $n$-gram  convolutions over the matrix. Here we incorporate context directly into the term 
encodings; hence similarities in this space are already context-sensitive. One way to view this difference is the point at which context enters the model -- 
directly during term encoding (Fig.~\ref{fig:ngram-term-reps}) or after term similarity scores have been computed (\pacrr, Fig.~\ref{fig:pacrr}).

\begin{figure}[t]
\includegraphics[width=3.2in]{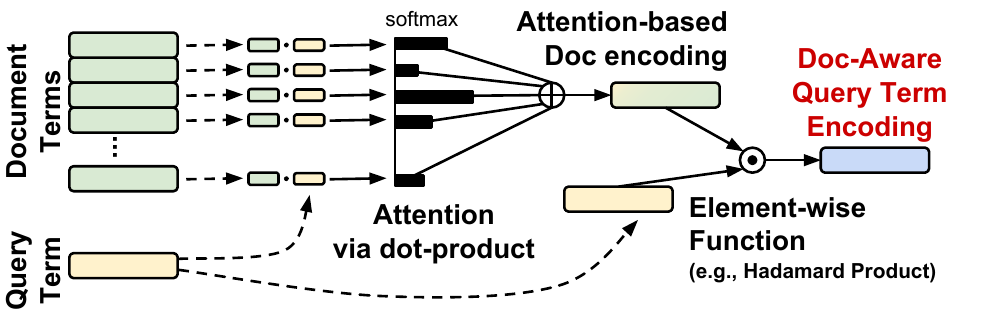}
\caption{\abeldrmm sub-net. From context-aware q-term and d-term encodings (Fig.~\ref{fig:ngram-term-reps}), it generates fixed-dimension \emph{document-aware q-term encodings} to be used in \drmm (Fig.~\ref{fig:top-level-drmm}, replacing $\otimes$ nodes).} 
\label{fig:hadamard-drmm}
\end{figure}

To make \drmm trainable end-to-end, we replace its histogram-based document-aware q-term encodings ($\otimes$ nodes of Fig.~\ref{fig:top-level-drmm}) by q-term encodings that consider d-terms via an attention-mechanism. Figure~\ref{fig:hadamard-drmm} shows the new sub-network that computes the document-aware encoding of a q-term $q_i$, given a document $d = \left<d_1, \ldots, d_m\right>$ of $m$ d-terms. We first compute a dot-product attention score $a_{i,j}$ for each $d_j$ relative to $q_i$:
\begin{equation}
a_{i,j} = \textrm{softmax}\Big(c(q_i)^T \, c(d_j); d_1, \dots, d_m\Big)
\label{eq:dpattention}
\end{equation}
where $c(t)$ is the context-sensitive encoding of $t$ (Eq.~\ref{eq:ct}). We then sum the context-sensitive encodings of the d-terms, weighted by their attention scores, to produce an attention-based representation $d_{q_i}$ of document $d$ from the viewpoint of $q_i$:
\begin{equation}
d_{q_i} = \sum_j a_{i,j} \; c(d_j)
\label{eq:dqi}
\end{equation}
The Hadamard product (element-wise multiplication, $\odot$) between 
the document representation $d_{q_i}$ and the q-term encoding $c(q_i)$ is then computed and used as the fixed-dimension document-aware encoding $\phi_{H}(q_i)$ of $q_i$ (Fig.~\ref{fig:hadamard-drmm}):
\begin{equation}
\phi_{H}(q_i) = d_{q_i}\odot c(q_i)
\label{eq:hadamard}
\end{equation}
The $\otimes$ nodes and lower parts of the \drmm network of Fig.~\ref{fig:top-level-drmm} are now replaced by (multiple copies of) the sub-network of Fig.~\ref{fig:hadamard-drmm} (one copy per q-term), with the $\odot$ nodes replacing the $\otimes$ nodes. We call the resulting model Attention-Based Element-wise \drmm (\abeldrmm).

Intuitively, if the document contains one or more terms $d_j$ that are similar to $q_i$, the attention mechanism will have emphasized mostly those terms and, hence, $d_{q_i}$ will be similar to $c(q_i)$, otherwise not. This similarity could have been measured by the cosine similarity between $d_{q_i}$ and $c(q_i)$, but the 
cosine similarity assigns the same weight to all the dimensions, i.e., to all the element-wise products in $\phi_{H}(q_i)$. By using the Hadamard product, we pass on to the upper layers of \drmm (the dense layers of Fig.~\ref{fig:top-level-drmm}), which score each q-term with respect to the document, all the element-wise products of $\phi_H(q_i)$, allowing the upper layers to learn which element-wise products (or combinations of them) are important when matching a q-term to the document.

\subsubsection{ABEL-DRMM extensions}
\label{sec:drmm-extensions}

We experimented with two extensions to \abeldrmm. The first is a \emph{density-based extension} that considers all the windows of $l_w$ consecutive tokens of the document and computes the \abeldrmm relevance score per window. The final relevance score of a document is the sum of the original \abeldrmm score computed over the entire document plus the maximum \abeldrmm score over all the document's windows. The intuition is to reward not only documents that match the query, but also those that match it in a dense window.

The second extension is to compute a \emph{confidence} score per document and only return documents with scores above a threshold. We apply a softmax over the \abeldrmm scores of the top $t_d$ documents and return only documents from the top $t_d$ with normalized scores  exceeding a threshold $t_c$. While this will always hurt metrics like Mean Average Precision (\textsc{map}) when evaluating document retrieval, it has the potential to improve the precision of downstream components, in our case snippet retrieval, which in fact we observe. 

\section{Snippet Retrieval}
\label{sec:snippets}

For the snippet retrieval task, we used the `basic \cnn' (\bcnn) network of the broader \abcnn model \cite{YinW2016bcnn}, which we combined with a post-processing stage, as discussed below. The input of snippet retrieval is an English question and text snippets (e.g., sentences) from documents that the document retrieval component returned as relevant to the question. The goal is to rank the snippets, so that snippets that human experts selected as relevant to the question will be ranked higher than others. In \bioasq, human experts are instructed to select relevant snippets consisting of one or more consecutive sentences.\footnote{This was not actually the case in \bioasq year 1. Hence, some of our training data do not adhere to this rule.} For training purposes, we split the relevant documents into sentences, and consider sentences that overlap the gold snippets (the ones selected by the human experts) as relevant snippets, and the remaining ones as irrelevant. At inference time, documents returned by the document retrieval model as relevant are split into sentences, and these sentences are ranked by the system. For details on sentence splitting, tokenization, etc., see Section~\ref{sec:resources}.

\subsection{BCNN Model}
\label{sec:BCNN}

\bcnn receives as input two sequences of terms (tokens), in our case a question (query) and a sentence from a document. All terms are represented by pre-trained  embeddings (Section~\ref{sec:resources}). Snippet sequences were truncated (or zero padded) to be of uniform length. A convolution layer with multiple filters, each of the same width $w$, is applied to each one of the two input sequences, followed by a windowed-average pooling layer over the same filter width  to produce a feature map (per filter) of the same dimensionality as the input to the convolution layer.\footnote{The same filters are applied to both queries and snippets.}
Consequently, we can stack an arbitrary number of convolution/pooling blocks in order to extract increasingly abstract features. 

\begin{figure}[t]
\centering
\includegraphics[width=3in]{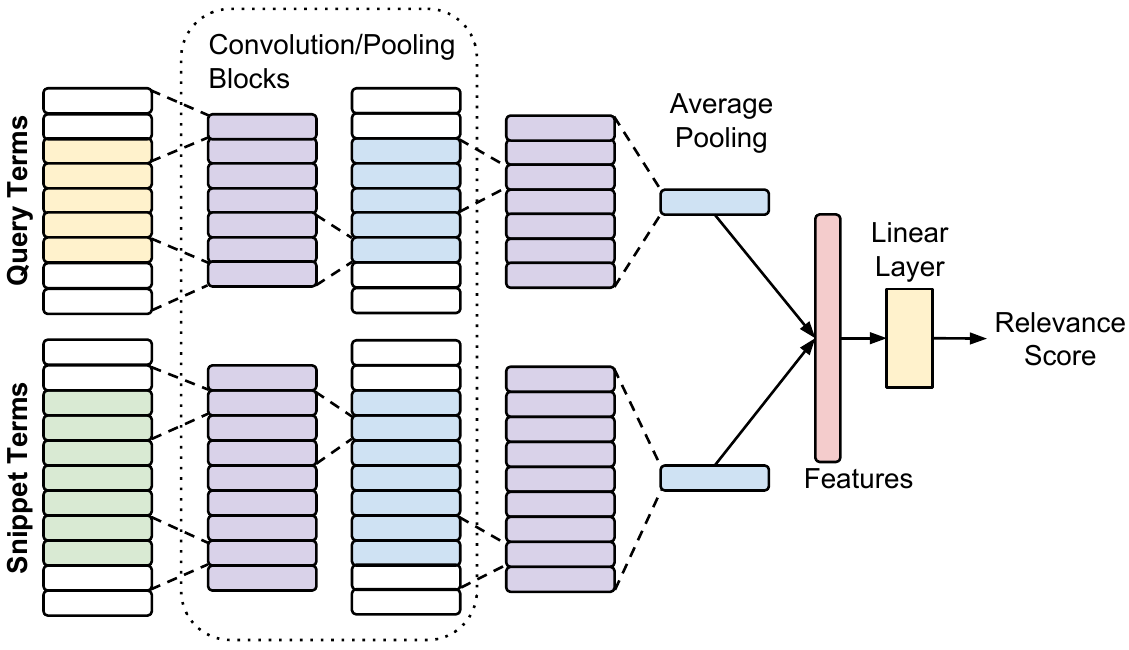}
\caption{\bcnn \cite{YinW2016bcnn} scoring snippets relative to a query. The example illustrates a query of 5 terms, a snippet of 7 terms, and a single convolution filter of width $w=3$. Zero-padding shown as empty boxes. In each convolution/pooling block, the convolution layer is followed by a windowed-average pooling of the same width $w$ to preserve the dimensionality of the input to the block. Thus convolution/pooling blocks can be repeated, making the model arbitrarily deep.} 
\label{fig:bcnn}
\end{figure}

An average pooling layer is then applied to the entire output of the last convolution/pooling block (Fig.~\ref{fig:bcnn}) to obtain a feature vector of the query and snippet, respectively. When multiple convolution filters are used (Fig.~\ref{fig:bcnn} illustrates only one), we obtain a different feature vector from each filter (for the query and snippet, respectively), and the feature vectors from the different filters are concatenated, again obtaining a single feature vector for the query and snippet, respectively. Similarity scores are then computed from the query and snippet feature vectors, and these are fed into a linear logistic regression layer. One critical implementation detail from the original \bcnn paper is that when computing the query-snippet similarity scores, average pooling is actually applied to the output of each one of the convolution/pooling blocks, i.e., we obtain a different query and snippet feature vector from the output of each block. Different similarity scores are computed based on the query and snippet feature vectors obtained from the output of each block, and all the similarity scores are passed to the final layer. Thus the number of inputs to the final layer is proportional to the number of blocks.

\subsection{Post-processing}
\label{sec:BCNNPostProcessing}

A technique that seems to improve our results in snippet retrieval is to retain only the top $K_s$ snippets with the best \bcnn scores for each query, and then re-rank the $K_s$ snippets by the relevance scores of the documents they came from; if two snippets came from the same document, they are subsequently ranked by their \bcnn score. This is a proxy for more sophisticated models that would jointly consider document and snippet retrieval. This is important as the snippet retrieval model is trained under the condition that it only sees relevant documents. So accounting for the rank/score of the document itself helps to correctly bias the snippet model.

\section{Overall System Architecture}
\label{sec:architecture}

\begin{figure*}[t]
\centering
\includegraphics[width=6in]{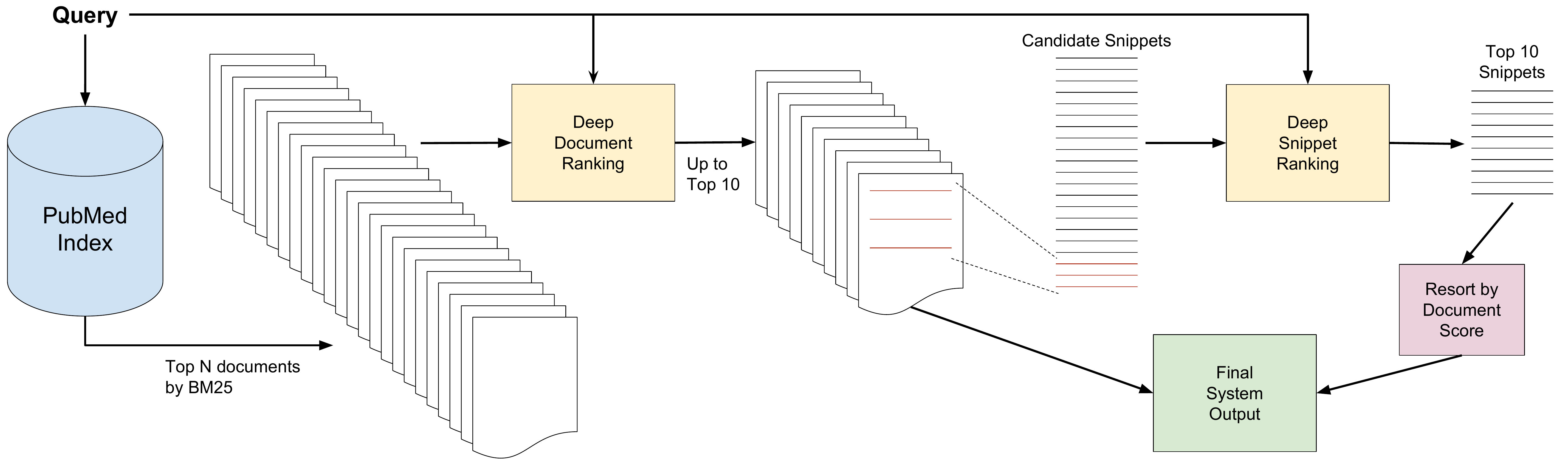}
\caption{Overall architecture of document and snippet retrieval systems.} 
\label{fig:architecture}
\end{figure*}

Figure~\ref{fig:architecture} outlines the general architecture that we used to piece together the various components. It consists of retrieving the top $N$ documents per query using \bmtf \cite{robertson1995bm25}; re-ranking the top $N$ documents using one of the document retrieval models (Section~\ref{sec:documentRetrieval}) and retaining (up to) the top $K_d$ documents; scoring all candidate snippets of the top $K_d$ documents via a snippet retrieval model (\bcnn, Section~\ref{sec:BCNN}) and retaining (up to) the top $K_s$ snippets; re-ranking the $K_s$ snippets by the relevance scores of the documents they came from (Section~\ref{sec:BCNNPostProcessing}).\footnote{The last step was used only in batches 3--5.}

We set $K_d = K_s =10$ as it was dictated by the \bioasq challenge. We set $N=100$ 
as we found that with this value, \bmtf returned the majority of the relevant documents from the training/development data sets. Setting $N$ to larger values had no impact on the final results. The reason for using a pre-retrieval model based on \bmtf is that the deep document retrieval models we use here are computationally expensive. Thus, running them on every document in the index for every query is prohibitive, whereas running them on the top $N=100$ documents from a pre-retrieval system is easily achieved.

\section{Experiments}
\label{sec:experiments}

All retrieval components (\pacrr-, \drmm-, \bcnn-based) were augmented to combine the scores of the corresponding deep model with a number of traditional \ir features, which is a common technique \cite{severyn2015learning}. In \termpacrr, the additional features are fed to the linear layer that combines the q-term scores (Fig.~\ref{fig:pacrr}). In \abeldrmm, an additional linear layer is used that concatenates the deep learning document relevance score with the traditional \ir features. In \bcnn, the additional features are included in the final linear layer (Fig.~\ref{fig:bcnn}). The additional features we used were the 
\bmtf score of the document (the document the snippet came from, in snippet retrieval), word overlap (binary and \idf weighted) between the query and the document or snippet; bigram overlap between the query and the document or snippet. The latter features were taken from \newcite{mohan2017deep}. The additional features improved the performance of all models.

\subsection{Data Resources and Pre-processing}
\label{sec:resources}

The document collection consists of approx.\
28M `articles' (titles and abstracts only) from the `\medline/PubMed Baseline 2018' collection.\footnote{Available from \url{https://www.nlm.nih.gov/databases/download/pubmed_medline.html}.} We discarded the approx.\ 10M articles that contained only titles, since very few of these were annotated as relevant. For the remaining 18M articles, a document was the concatenation of each title and abstract. These documents were then indexed using Galago, removing stop words and applying Krovetz's stemmer \cite{krovetz1993krovetzstemmer}.\footnote{We used Galago version 3.10. Consult \url{http://www.lemurproject.org/galago.php}.} This served as our pre-retrieval model.

Word embeddings were pre-trained by applying word2vec \cite{mikolov2013distributed} to the 28M `articles' of the \medline/PubMed collection. \idf values were computed over the 18M articles that contained both titles and abstracts. We used the GenSim implementation of word2vec (skip-gram model), with negative sampling, window size set to 5, default other hyper-parameter values, to produce word embeddings of 200 dimensions.\footnote{Consult \url{https://radimrehurek.com/gensim/models/word2vec.html}. We used Gensim v.~3.3.0. The word embeddings and code of our experiments are available at 
\url{https://github.com/nlpaueb/aueb-bioasq6}.
}
The word embeddings were not updated when training the document relevance ranking models. For tokenization, we used the `bioclean' tool provided by \bioasq.\footnote{The tool accompanies an older set of embeddings provided by \bioasq. See \url{http://participants-area.bioasq.org/tools/BioASQword2vec/}.} 
In snippet retrieval, we used \textsc{nltk}'s English sentence splitter.\footnote{We used \textsc{nltk} v3.2.3. See \url{https://www.nltk.org/api/nltk.tokenize.html}.}

To train and tune the models we used years 1--5 of the \bioasq data, using batch 5 of year 5 as development for the final submitted models, specifically when selecting optimal model epoch. We report test results (\textsc{f1}, \map, \gmap) on batches 1--5 of year 6 from the official results table.\footnote{Available at \url{http://participants-area.bioasq.org/results/6b/phaseA/}. The names of our systems have been modified for the blind review.} Details on the three evaluation metrics are provided by \newcite{tsatsaronis2015overview}. They are standard, with the exception that \map here always assumes 10 relevant documents/snippets, which is the maximum number of documents/snippets the participating systems were allowed to return per query.

\subsection{Hyperparameters}
\label{sec:hyperparameters}

All \drmm-based models were trained with Adam \cite{kingma2014adam} with a learning rate of 0.01 and $\beta_1/\beta_2=0.9/0.999$. Batch sizes were set to 32. We used a hinge-loss with a margin of 1.0 over pairs of a single positive and a single negative document of the same query. All models used a two-layer \mlp to score q-terms (dense layers of Fig.~\ref{fig:top-level-drmm}), with leaky-\relu activation functions and 8 dimensions per hidden layer. For context-sensitive term encodings (Fig.~\ref{fig:ngram-term-reps}), a single layer was used, again with leaky-\relu as activation. For the density-based extension of \abeldrmm (Section~\ref{sec:drmm-extensions}), $l_w =20$. For the confidence extension of \abeldrmm, $t_d =100$, $t_c = 0.01$.

\termpacrr was also trained with Adam, with a learning rate of 0.001 and $\beta_1/\beta_2=0.9/0.999$ with batch size equal to 32. Following \newcite{hui2018copacrr}, we used binary log-loss over pairs of a single positive and a single negative document of the same query.
Maximum query length $l_q$ was set to 30 and maximum document length $l_d$ was set to 300. Maximum kernel size $(l_g \times l_g)$ was set to $(3 \times 3)$ with 16 filters per size. Row-wise $k$-max pooling used $k=2$. \termpacrr used a two-layer \mlp with \relu activations and hidden layers with 7 dimensions to independently score each document-aware query-term encoding. 

\bcnn was trained using binary log-loss and AdaGrad \cite{DuchiAdagrad}, with a learning rate of 0.08 and $L2$ regularization with $\lambda = 0.0004$. We used 50 convolution kernels (filters) of width $w$ = 4 in each convolution layer, and two convolution/pooling blocks. Finally, batch sizes were set to 200. Snippets were truncated to 40 tokens. Questions were never truncated.

\subsection{Official Submissions}
\label{sec:official}

We submitted 5 different systems to the BioASQ challenge, all of which consist of components described above.

\begin{itemize}
\item \textbf{\auebnlp-1}: Combo of 10 runs of \termpacrr for document retrieval ($\S$\ref{sec:pacrr-bioasq}) followed by \bcnn for snippet retrieval ($\S$\ref{sec:snippets}).
\item \textbf{\auebnlp-2}: Combo of 10 runs of \abeldrmm ($\S$\ref{sec:drmm-bioasq}) for document retrieval followed by \bcnn for snippet retrieval.
\item \textbf{\auebnlp-3}: Combo of 10 runs of \termpacrr and 10 runs of \abeldrmm followed by \bcnn for snippet retrieval.
\item \textbf{\auebnlp-4}: \abeldrmm with density extension ($\S$\ref{sec:drmm-extensions}) for document retrieval followed by \bcnn for snippet retrieval.
\item \textbf{\auebnlp-5}: 
\abeldrmm with both density and confidence extensions ($\S$\ref{sec:drmm-extensions}) for document retrieval followed by \bcnn for snippet retrieval.
This system was submitted for batches 2-5 only.
\end{itemize}

In combination (combo) systems, we obtained 10 versions of the corresponding model by retraining it 10 times with different random seeds, and we then used a simple voting scheme. If a document was ranked at position 1 by a model it got 10 votes, position 2 was 9 votes, until position 10 where it got 1 vote. Votes were then aggregated over all  models in the combination. While voting did not improve upon the best single model, it made the results more stable across different runs.

\subsection{Results}
\label{sec:results}

Results are given in Table~\ref{tab:bioasq-6b}. There are a number of things to note. First, for document retrieval, there is very little difference between our submitted models. Both \pacrr- and \drmm-based models perform well (usually at the top or near the top) with less than 1 \map point separating them. These systems were all competitive and for 4 of the 5 batches one was the top scoring system in the competition. On average the experimental \abeldrmm system (\auebnlp-4) scored best amongst \aueb submissions and in aggregate over all submissions, but by a small margin (0.1053 average \map versus 0.1016 for \termpacrr). The exception was the high precision system (\auebnlp-5) which did worse in all metrics except F1, where it was easily the best system for the 4 batches it participated in. This is not particularly surprising, but impacted snippet selection, as we will see.

For snippet selection, all systems did well (\auebnlp-[1-4]) and it is hard to form a pattern that a base document retrieval model's results are more conducive to snippet selection. The exception is the high-precision document retrieval model of \auebnlp-5, which had by far the best scores for \aueb submissions and the challenge as a whole. The main reason for this is that the snippet retrieval component was trained assuming only relevant documents as input. Thus, if we fed it all 10 documents, even when some were not relevant, it could theoretically still rank a snippet from an irrelevant document high since it is not trained to combat this. By sending the snippet retrieval model only high precision document sets it focused on finding good snippets at the expense of potentially missing some relevant documents.

\begin{table*}[t]
\begin{minipage}{3in}
\centering
\small
{\bf DOCUMENT RETRIEVAL}\vspace{0.01in}
\begin{tabular}{|l|ccc|}
\hline
{\bf System}& {\bf F1} & {\bf MAP}  & {\bf GMAP}   \\ \hline
\multicolumn{4}{|c|}{Batch 1} \\ \hline
\auebnlp-1 & 0.2546 & 0.1246 & 0.0282\\
\auebnlp-2 & 0.2462 & 0.1229 & 0.0293\\
\textit{\textbf{\auebnlp-3}} & 0.2564 & 0.1271 & 0.0280\\
\auebnlp-4 & 0.2515 & 0.1255 & 0.0235\\ \hline
Top Competitor & 0.2216 & 0.1058 & 0.0113 \\ \hline
\multicolumn{4}{|c|}{Batch 2} \\ \hline
\auebnlp-1 & 0.2264 & 0.1096 & 0.0148 \\
\textit{\textbf{\auebnlp-2}} & 0.2473 & 0.1207 & 0.0200 \\
\auebnlp-3 & 0.2364 & 0.1178 & 0.0161 \\
\auebnlp-4 & 0.2350 & 0.1182 & 0.0161 \\
\auebnlp-5 & 0.3609 & 0.1014 & 0.0112 \\ \hline
Top Competitor & 0.2265 & 0.1201 & 0.0183 \\ \hline
\multicolumn{4}{|c|}{Batch 3} \\ \hline
\auebnlp-1 & 0.2345 & 0.1122 & 0.0101 \\
\textit{\auebnlp-2} & 0.2345 & 0.1147 & 0.0108 \\
\auebnlp-3 & 0.2350 & 0.1135 & 0.0109 \\
\auebnlp-4 & 0.2345 & 0.1137 & 0.0106 \\
\auebnlp-5 & 0.4093 & 0.0973 & 0.0062 \\ \hline
Top Competitor & 0.2186 & 0.1281 & 0.0113 \\ \hline
\multicolumn{4}{|c|}{Batch 4} \\ \hline
\auebnlp-1 & 0.2136 & 0.0971 & 0.0070 \\
\auebnlp-2 & 0.2148 & 0.0996 & 0.0069 \\
\textit{\textbf{\auebnlp-3}} & 0.2134 & 0.1000 & 0.0068 \\
\auebnlp-4 & 0.2094 & 0.0995 & 0.0064 \\
\auebnlp-5 & 0.3509 & 0.0875 & 0.0044 \\ \hline
Top Competitor & 0.2044 & 0.0967 & 0.0073 \\ \hline
\multicolumn{4}{|c|}{Batch 5} \\ \hline
\auebnlp-1 & 0.1541 & 0.0646 & 0.0009 \\
\auebnlp-2 & 0.1522 & 0.0678 & 0.0013 \\
\auebnlp-3 & 0.1513 & 0.0663 & 0.0010 \\
\textit{\textbf{\auebnlp-4}} & 0.1590 & 0.0695 & 0.0012 \\
\auebnlp-5 & 0.1780 & 0.0594 & 0.0008 \\ \hline
Top Competitor & 0.1513 & 0.0680 & 0.0009 \\ \hline
\end{tabular}
\end{minipage}
\begin{minipage}{3in}
\centering
\small
{\bf SNIPPET RETRIEVAL}\vspace{0.01in}
\begin{tabular}{|l|ccc|}
\hline
{\bf System} & {\bf F1} & {\bf MAP} & {\bf GMAP} \\ \hline
\multicolumn{4}{|c|}{Batch 1} \\ \hline
\auebnlp-1 & 0.1296 & 0.0687 & 0.0029 \\
\auebnlp-2 & 0.1347 & 0.0665 & 0.0026 \\
\auebnlp-3 & 0.1329 & 0.0661 & 0.0028 \\
\textit{\auebnlp-4} & 0.1297 & 0.0694 & 0.0024 \\ \hline
Top Competitor & 0.1028 & 0.0710 & 0.0002 \\ \hline
\multicolumn{4}{|c|}{Batch 2} \\ \hline
\auebnlp-1 & 0.1329 & 0.0717 & 0.0034 \\
\auebnlp-2 & 0.1434 & 0.0750 & 0.0044 \\
\auebnlp-3 & 0.1355 & 0.0734 & 0.0033 \\
\auebnlp-4 & 0.1397 & 0.0713 & 0.0037 \\
\textit{\textbf{\auebnlp-5}} & 0.1939 & 0.1368 & 0.0045 \\ \hline
Top Competitor & 0.1416 & 0.0938 & 0.0011 \\ \hline
\multicolumn{4}{|c|}{Batch 3} \\ \hline
\auebnlp-1 & 0.1563 & 0.1331 & 0.0046 \\
\auebnlp-2 & 0.1494 & 0.1262 & 0.0034 \\
\auebnlp-3 & 0.1526 & 0.1294 & 0.0038 \\
\auebnlp-4 & 0.1519 & 0.1293 & 0.0038 \\
\textit{\textbf{\auebnlp-5}} & 0.2744 & 0.2314 & 0.0068 \\ \hline
Top Competitor & 0.1877 & 0.1344 & 0.0014 \\ \hline
\multicolumn{4}{|c|}{Batch 4} \\ \hline
\auebnlp-1 & 0.1211 & 0.0716 & 0.0009 \\
\auebnlp-2 & 0.1307 & 0.0821 & 0.0011 \\
\auebnlp-3 & 0.1251 & 0.0747 & 0.0009 \\
\auebnlp-4 & 0.1180 & 0.0750 & 0.0009 \\
\textit{\textbf{\auebnlp-5}} & 0.1940 & 0.1425 & 0.0017 \\ \hline
Top Competitor & 0.1306 & 0.0980 & 0.0006 \\ \hline
\multicolumn{4}{|c|}{Batch 5} \\ \hline
\auebnlp-1 & 0.0768 & 0.0357 & 0.0003 \\
\auebnlp-2 & 0.0728 & 0.0405 & 0.0004 \\
\auebnlp-3 & 0.0747 & 0.0377 & 0.0004 \\
\auebnlp-4 & 0.0790 & 0.0403 & 0.0004 \\
\textit{\textbf{\auebnlp-5}} & 0.0778 & 0.0526 & 0.0003 \\ \hline
Top Competitor & 0.0542 & 0.0475 & 0.0001 \\ \hline
\end{tabular}
\end{minipage}
\caption{Performance on \bioasq Task 6b, Phase A (batches 1--5) for document and snippet retrieval (left and right tables, respectively). Systems described in Section~\ref{sec:official}. The italicised system is the top scoring system from \aueb's entries and if also in bold, is the top from all official entries in that batch. \emph{Top} is by \map, the official metric of \bioasq. \emph{Top Competitor} is the top scoring entry -- by \map -- that is not among \aueb's submissions.}
\label{tab:bioasq-6b}
\end{table*}

\section{Related Work}

Document ranking has been studied since the dawn of \ir; classic term-weighting schemes were designed for this problem \cite{sparck1972statistical,robertson1976relevance}. With the advent of statistical \nlp and statistical \ir, probabilistic language and topic modeling were explored \cite{zhai2001study,wei2006lda}, followed  recently by deep learning \ir methods \cite{lu2013deep,hu2014convolutional,palangi2016deep,guo2016deep,hui2017pacrr}.

Most document relevance ranking methods fall within two categories: representation-based, e.g., \newcite{palangi2016deep}, or interaction-based, e.g., \newcite{lu2013deep}. In the former, representations of the query and document are generated independently. Interaction between the two only happens at the final stage, where a score is generated indicating relevance. End-to-end learning and backpropagation through the network tie the two representations together. In the interaction-based paradigm -- which is where the models studied here fall -- 
explicit encodings between pairs of queries and documents are induced. This allows direct modeling of exact- or near-matching terms (e.g., synonyms), which is crucial for relevance ranking.
Indeed, \newcite{guo2016deep} showed that the interaction-based \drmm  outperforms previous representation-based methods. 
On the other hand, interaction-based models are less efficient, since one cannot index a document representation independently of the query. This is less important, though, when relevance ranking methods rerank the top documents returned by a conventional \ir engine, which is the scenario we consider here.

In terms of biomedical document and snippet retrieval, several methods have been proposed for \bioasq \cite{tsatsaronis2015overview}, mostly based on traditional \ir and \ml techniques. For example, the system of \newcite{jin2017multi}, which is the top scoring one for previous incarnations of \bioasq (\textsc{utsb} team),  uses an underlying graphical model for scoring coupled with a number of traditional \ir techniques like pseudo-relevance feedback.

The most related work from the biomedical domain is that of \newcite{mohan2017deep}, who use a deep learning architecture for document ranking. Like our systems they use interaction-based models to score and aggregate q-term matches relative to a document, 
however using different document-aware q-term representations -- namely best match d-term distance scores. Also unlike our work, they focus on user click data as a supervised signal, and they use context-insensitive representations of document-query term interactions.

There are several studies on deep learning systems for snippet selection which aim to improve the classification and ranking of snippets extracted from a document based on a specific query. \newcite{WangBiLSTMs} use a stacked bidirectional \lstm (\bilstm); their system gets as input a question and a sentence, it concatenates them in a single string and then forwards that string to the input layer of the \bilstm. \newcite{RaoNoiseConstractive} employ a neural architecture to produce representations of pairs of the form (\textit{question}, \textit{sentence}) and to learn to rank pairs of the form (\textit{question}, \textit{relevant sentence}) higher than pairs of the form (\textit{question}, \textit{irrelevant sentence}) using Noise-Contrastive Estimation. Finally, \newcite{AmiriAutoencoders} use autoencoders to learn to encode input texts and use the resulting encodings to compute similarity between text pairs. This is similar in nature to \bcnn, the main difference being  the encoding mechanism.

\section{Conclusions}

We presented the models, experimental set-up, and results of \aueb's submissions to the document and snippet retrieval tasks of the sixth year of the \bioasq challenge. Our results show that deep learning models are not only competitive in both tasks, but in aggregate were the top scoring systems. This is in contrast to previous years where traditional \ir systems tended to dominate. In future years, as deep ranking models improve and training data sets get larger, we expect to see bigger gains from deep learning models.



\bibliography{emnlp2018}
\bibliographystyle{acl_natbib_nourl}
\end{document}